\documentstyle[preprint,aps]{revtex}

\begin{document}

\tightenlines

\draft

\title{Stochastic Motion of an Open Bosonic String}

\author{L. F. Santos $^1$ and C. O. Escobar $^2$}

\address{$^1$Instituto de F\'{\i}sica da Universidade de S\~ao Paulo, 
C.P. 66318, cep 05389-970 \\
S\~ao Paulo, S\~ao Paulo, Brazil\\
lsantos@charme.if.usp.br \\
$^2$ Departamento de  Raios C\'osmicos e Cronologia \\
Instituto de F\'{\i}sica Gleb Wataghin \\
Universidade Estadual de Campinas, C.P. 6165,  cep 13083-970\\
Campinas, S\~ao Paulo, Brazil\\
escobar@ifi.unicamp.br}

\maketitle

\begin{abstract}
We show that the classical stochastic motion of an open bosonic string leads 
to the same results as the standard first quantization of this system. 
For this, the diffusion constant governing the process has to be proportional 
to $\alpha' $, the Regge slope parameter, 
which is the only constant, along with the velocity 
of light, needed to describe the motion 
of a string.
\end{abstract}


\section{Introduction}

String theory \cite{Green} provides an example of the not always 
straight path towards new ideas in physics. From its inception in the 
framework of dual-resonance models \cite{Nambu} to its current position 
as a candidate for a unified theory of all interactions, including 
gravity, it has gone through several modifications and increasing 
sophistication in its mathematical formulation. As a consequence of 
these changes, the original motivation for strings was abandoned and 
what we have now is a description of physics at the Planck scale 
\cite{Veneziano}, bringing 
with it all uncertainties on the validity of 
the basic quantum mechanical ideas. For sure we lack a clear physical 
picture of phenomena at such small distances, despite recent efforts by 
several authors \cite{Susskind} 
who introduce a radical view of physics 
at this scale, expanding the traditional quantum mechanical view. Some 
authors have gone as far as to speculate that at the Planck scale, 
quantum and thermal fluctuations cannot be distinguished \cite{Smolin}.

Motivated by the above considerations we decided to look at a classical 
string subjected to stochastic motion, very much in the spirit of 
Nelson's stochastic approach to the motion of a classical newtonian 
particle \cite{Nelson}. This formulation achieves 
the derivation of a Sch\"odinger 
equation for a classical
non-relativistic particle moving in an external potential by a 
stochastic version of Newton's second law of motion. There have been 
several criticisms to Nelson's alternative to quantum mechanics
\cite{Wallstrom,Davidson} and we 
do not want to address neither the criticisms nor 
the arguments purporting to defend it \cite{book}.
However, we think that if stochastic fluctuations, whose nature are 
never specified by Nelson and his followers, have any chance of 
manifesting themselves, then the natural place for them to occur would 
be at the very small scales like the Planck scale \cite{Smolin}. Since 
the objects that are candidates to describe physics at this scale are 
strings, we will in the following describe the stochastic motion of an 
open bosonic string, as the simplest such an object.

This paper is organized as follows. In the next section we briefly review 
Nelson's stochastic mechanics and very succinctly describe the stochastic 
variational principle, developed by Guerra and his coworkers 
\cite{Morato}, which will be useful in section III, where our stochastic 
formulation of an open bosonic string is presented. Among the results 
obtained, we can mention the derivation of a wave-functional equation 
for the string, the existence of a critical dimension ($D = 26$), 
obtained through the requirement of Lorentz invariance (as we 
work in the light-cone gauge) and the two-point correlation function 
for the non-zero normal modes of the string. Our 
conclusions are in section IV.

\section{Stochastic Mechanics}

\subsection{Nelson's approach}

The starting point of Nelsons's approach is to consider the stochastic 
motion of a point particle (for simplicity we treat a one dimensional 
motion) given by

\begin{equation}
dq(x,t) = v_{+} (q(t), t) \, dt + dw(t) \; ,
\end{equation}

where the first term on the right-hand side is deterministic and
introduces a velocity field for forward propagation ($dt>0$) written as

\begin{equation}
v_{+} (x,t)= \frac{\nabla S_{+} (x,t)}{m} \; ,
\end{equation}

with $m$ the particle mass and $S_{+}$ a scalar function. The stochastic 
process is described by $dw(t)$, which satisfies the following averages

\begin{eqnarray}
<dw(t)>     & = & 0 \\
<dw(t)\, dw(t)> & = & 2 \nu dt \; .
\end{eqnarray}

In (4) $\nu $ is a diffusion constant to be specified later.

Given the non-differentiable nature of (1), Nelson then introduces the 
mean backward and forward transport derivatives

\begin{eqnarray}
(D_{+} q)(x,t) &=& \lim _{\Delta t \rightarrow 0^{+}}
\frac{<q(t + \Delta
t) - q(t)>}{\Delta t} = v_{+} (x,t) \\
(D_{-} q)(x,t) &=& \lim _{\Delta t \rightarrow 0^{+}}\frac{<q(t) - q(t - 
\Delta t)>}{\Delta t} = v_{-} (x,t) \; ,
\end{eqnarray}

which for a function $F$ of $x$ and $t$ can be written, using (1) and 
(4) as

\begin{equation}
(D_{\pm} F)(x,t) = (\partial _{t} F)(x,t) + v_{\pm} (x,t) (\nabla
F)(x,t) \pm \nu (\nabla ^2 F)(x,t) \; .
\end {equation}

(1) and (4) also imply the Fokker-Planck equation
   
\begin{equation}
\partial _{t} p(x,t;x_0 ,t_0 ) = -\nabla  (v_{+} (x,t) p(x,t;x_0 ,t_0 )) 
+ \nu 
\nabla ^2 p(x,t;x_0 ,t_0 ) \; ,
\end{equation}

where $p$ is the transition probability, which, by definition, propagates 
the probability density of an ensemble of particles

\begin{equation}
\rho (x,t) = \int p(x,t; x_0 ,t_0 ) \rho (x_0 , t_0 ) dx_0 \; .
\end{equation}

With $v_{+}$ and $v_{-}$ we can also define

\begin{eqnarray}
v &=& \frac{1}{2} (v_{+} + v_{-}) = \frac{\nabla S}{m} \\
u &=& \frac{1}{2} (v_{+} - v_{-}) = \nu \frac{\nabla \rho}{\rho }
\end{eqnarray}

and then obtain the continuity equation

\begin{equation}
\partial _{t} \rho(x,t) = -\nabla  (\rho (x,t) v(x,t) ) \; .
\end{equation}

Nelson's formulation of the second law is \cite{Davidson}

\begin{equation}
\frac{1}{2} [(D_{+} D_{-} + D_{-} D_{+} )q](x,t) = -\frac{1}{m} (\nabla 
V)(x) \; .
\end{equation}

From this equation of motion follows a Madelung type equation

\begin{equation}
\partial _{t} S + \frac{(\nabla  S)^2 }{2m} - 2m \nu ^2 [(\nabla R)^2 + 
\nabla ^2 R ]= -V(x) \; ,
\end{equation}

where $R$ is related to the probability density $\rho $ as follows

\begin{equation}
\exp (2R(x,t)) = \rho (x,t) \; .
\end{equation}

The Madelung equation and the continuity equation will correspond 
respectively to the real and imaginary parts of a Schr\"odinger equation 
when writing the wave function in polar form

\begin{equation}
\psi (x,t) = \exp (R(x,t)) \exp \left(\imath \frac{S(x,t)}{\hbar }\right)
\end{equation}

provided the diffusion constant $\nu $ is identified with $\frac{\hbar 
}{2m}$ \cite{foot1,Davidson}.

\subsection{Stochastic Variational Principle}

In order to avoid the ambiguity of defining the acceleration in Newton's 
second law, Guerra and collaborators \cite{Morato} formulated stochastic 
mechanics with a variational principle. We refer the reader to ref.
\cite{Morato}
for more details, since here we only need their lagrangian density which 
will lead to a Madelung equation.

Starting from

\begin{equation}
L(x,t) = \frac{1}{2} m v_{+} (x,t) v_{-} (x,t) - V(x)
\end{equation}

and defining an average stochastic action

\begin{equation}
<A(t_{0} ,t; \rho _{0}; v_{+})> = \int _{t_{0} }^{t} \int L(x,t) \rho 
(x,t) dx\, dt
\end{equation}

where $\rho _{0} $ is the initial distribution, another lagrangian 
density depending only on $v_{+} $ can be introduced

\begin{equation}
L_{+} (x,t) = \frac{1}{2} m v_{+} ^2 (x,t) + m \nu (\nabla  v_{+} )
(x,t)- V(x) \; .
\end{equation}

It is possible to show that the action in equation (18) is the same as 
would be obtained replacing $L$ by $L_{+} $ (the extra terms in (19) 
vanish when taking the stochastic average).

Using a smooth field $S_{1} $ and $B(t_{0}, t; \rho _{0} , 
S_{1} ; v_{+})$ defined as

\begin{equation}
B(t_{0}, t; \rho _{0} , S_{1} ; v_{+} ) = A(t_{0} ,t; \rho _{0} ; v_{+} ) 
- <S_{1} (q( t_{1} ))> = - \int J( x_0 , t_0 ; t_1, S_1 ; v_{+} )
\rho _{0} ( x_0 ) d x_0
\end{equation}

we obtain

\begin{equation}
(D_{+} J )(x,t) = L_{+} (x,t)
\end{equation}

The variational principle based on $\delta B = 0$ gives $v(x,t) =
(\nabla S)(x,t) /m $ (where $S$ is $J$ making $B$ stationary) 
and allows 
the identification of (21) with the Madelung equation.

The continuity equation is the same as before and with the above 
Madelung equation, thus reproduce the Schr\"odinger equation (imaginary 
and real parts respectively). Furthermore, the variational approach 
leads naturally to a canonical stochastic formulation with $\rho $ and 
$S$ as canonical variables and, as shown by Guerra and Marra 
\cite{Marra}, to a redefinition of Poisson brackets in this context. We 
will make use of this formalism at the end of section III.

A final remark is important before we proceed to the stochastic 
description of the open bosonic string. In this section we summarized the 
stochastic approach to the non-relativistic point particle. In dealing 
with a string we should use a relativistic generalization of the Markov 
process underlying Nelson's stochastic formalism. This has been 
achieved in the literature \cite{Dohrn} and in the following we will use 
this generalization.

\section{Stochastic Motion of a String}

We start from the Nambu-Goto action for a string

\begin{equation}
S = - \frac{1}{2 \pi \alpha '} \int _{\tau _0 } ^{\tau _1 } d\tau
\int _{ 0 } ^{\pi }d\sigma \sqrt{ -\left( \frac{\partial x}{\partial \tau}
\right )^2 
\left( \frac{ \partial x}{\partial \sigma }\right )^2 + 
\left( \frac{\partial x }{\partial 
\tau} \frac{\partial x }{\partial \sigma }\right) ^2} \; ,
\end{equation}

which is proportional to the area of the 
two dimensional surface (embedded in a D 
dimensional space-time) swept by a string. The space-time points 
$x^{\mu } (\sigma, \tau )$ on this surface are labelled by $\sigma $ and 
$\tau $, two dimensionless parameters. $\alpha '$ 
is the Regge slope which in string theory 
is related to the square of a fundamental length \cite{Veneziano}. 

The above action is invariant under reparametrizations of the surface: 
$\tilde{\sigma } \rightarrow \sigma =\sigma (\tilde{\sigma }, 
\tilde{\tau })$, $\tilde{\tau } \rightarrow \tau =\tau 
(\tilde{\sigma }, \tilde{\tau })$, which introduces a gauge freedom into the 
theory.. 
This leads to difficulties in the 
canonical formalism already at the classical level and one is forced to 
fix the gauge before proceeding. We choose to work in the light-cone 
gauge \cite{Rebbi} in our classical approach. This choice 
of gauge will make Lorentz invariance not manifest and this will have to 
be addressed by the stochastic approach.

In the light-cone system one introduces coordinates as follows,

\begin{equation}
x^{+} = \frac{x^0 + x^{D-1}}{\sqrt{2}}
\end{equation}

\begin{equation}
x^{-} = \frac{x^0 - x^{D-1}}{\sqrt{2}}
\end{equation}

and fixes the gauge as

\begin{equation}
x^{+} (\sigma, \tau ) = p^{+} \tau \; .
\end{equation}

The transverse coordinates, the physical degrees of freedom, satisfy the 
following equation of motion,

\begin{equation}
\ddot{x_{i} } - x_{i }  ''= 0 \; \; (i =1,... ,D-2)
\end{equation}

where $\dot{x} = \frac{\partial x}{\partial \tau }$ and $x' =
\frac{\partial x}{\partial \sigma }$.

We can now approach the system described by (26) from a stochastic point 
of view. We do this by expanding $x_{i} (\sigma, \tau )$ in normal modes

\begin{equation}
x_{i} = \sum _{n=0} ^{\infty } 
x_{ni} \, \cos n\sigma
\end{equation}

(we are following the standard convention of defining $\sigma $ between 
$0$ and $\pi $). From which follows,

\begin{equation}
\ddot{x_{ni} } + n^2 x_{ni} = 0 \; .
\end{equation}

We now promote the normal modes to a stochastic process \cite{Guerra} 
$q_{n}^{i} [x_{m}^{i}, \tau ]$ with $m = 0,... , \infty $ and $i =
1,... , D-2 $, satisfying

\begin{equation}
d q_{n}^{i} [x_{m}^{i}, \tau ] = v_{+n}^{i} d\tau + d w_{n}^{i} \; ,
\end{equation}

where $d w_{n}^{i} $ obeys,

\begin{eqnarray}
<dw_{n}^{i} (\tau )>                       &=& 0 \\
<dw_{n}^{i} (\tau ) \; dw_{n'}^{j} (\tau)> &=& 2 \; 
\delta _{ij} \; \delta _{nn'}
\; \nu 
_{n} d\tau \; .
\end{eqnarray}

Equation (29) illustrates the use of $\tau $ as an evolution parameter.

Notice that we have not specified the diffusion constant $\nu _{n} $ and 
have also allowed, for the sake of generality, a possible dependence on 
the normal mode index.

Following the steps outlined in section II we derive an equation of 
continuity

\begin{equation}
\partial _{\tau } \rho[x_{m}^{i}, \tau ] = -\sum _{n} \nabla _{n}^{i} (\rho 
[x_{m}^{i}, \tau ] v_{n}^{i} [x_{m}^{i}, \tau ])
\end{equation}

where
\begin{equation}
v_{n}^{i} [x_{m}^{i}, \tau ]  = 4\alpha ' \, \nabla _{n}^{i} S [x_{m}^{i}, 
\tau ]
\end{equation}

and for the zero mode

\begin{equation}
v_{0}^{i} [x_{m}^{i}, \tau ] = 2\alpha '  \, \nabla _{0}^{i} S [x_{m}^{i}, 
\tau ] \; .
\end{equation}

We can introduce, as before, the transport derivatives, which now read

\begin{equation}
D_{\pm } = \partial _{\tau } + \sum _{n} v_{\pm n}^{i} \nabla _{n}^{i} 
\pm  \sum _{n}  (\nabla _{n}^{i})^2
\end{equation}

In the classical equation for the normal mode amplitudes

\begin{equation}
\ddot{q_{n}^{i}}  + n^2 q_{n}^{i} = 0
\end{equation}

we use the definition of stochastic acceleration

\begin{equation}
\frac{1}{2} (D_{+} D_{-} + D_{+} D_{-} ) q_{n}^{i} = -n^2  q_{n}^{i} \; .
\end{equation}

The Madelung equation that follows is

\begin{eqnarray}
\partial _{\tau } S &=& -\alpha ' ( \nabla _{0}^{i} S)^2 + 
\frac{\nu _{0} \nu_{n}}{2\alpha ' }[
( \nabla _{0}^{i} R)^2 + (\nabla _{0}^{i})^2 R] -
\sum _{m \neq 0} 2\alpha ' ( \nabla _{m}^{i} S)^2 + \\ \nonumber 
                   &+& \frac{\nu _{n} \nu_{n}}{2\alpha ' }
 \sum _{m \neq 0}
[( \nabla _{m}^{i} R)^2 + (\nabla _{m}^{i})^2 R] -
\frac{1}{4\alpha '} \sum _{m}  \frac{m^2 (x_{m}^{i})^2}{2} = 0 \; .
\end{eqnarray}

It satisfies, together with (32), the  wave 
functional equation for a string (using $\psi = \exp (R) \exp (\imath S) $)

\begin{equation}
\imath \partial _{\tau } \psi = \left[ -\alpha ' (\nabla _{0}^{i})^2 + 
\sum _{n\neq 0} (-2\alpha ' (\nabla _{n}^{i})^2 + 
\frac{n^2 (x_{n}^{i})^2}{8\alpha '}  )\right] \psi  \; ,
\end{equation}

provided the diffusion constants are

\begin{equation}
\nu _{n} = 2\alpha '   \; \;      n\neq 0 
\end{equation}

\begin{equation}
\nu _{0} = \alpha ' \; .
\end{equation}

The difference between $\nu _{n}$ and $\nu _{0} $
comes from the convention used for separating the zero mode.

Notice the fact that we started this analysis without knowing 
the diffusion constant governing the stochastic process (31). For 
consistency reasons it results that in the case of the 
stochastic motion of a 
string the diffusion constant is $\alpha '$, which accords with the 
point of view stressed by Veneziano that a stringy world 
has only two constants $c$ and $\lambda $, which 
is related to $\alpha '$ as $\lambda ^{2} = 2 \alpha ' $ \cite{Veneziano}. 
If we naively expected $\nu $ to be 
related to the energy of the system, like in  point particle mechanics, $\nu $
would be different for each state of the string. This  
encourages speculations on the indistinguishability of 
quantum and classical fluctuations at the Planck scale.

It follows from (38) that we obtain the standard spectrum of an open 
bosonic string, given by an infinite set of harmonic oscillators, as
expected.

\subsection{Lorentz Invariance}

We must now address the question of Lorentz invariance. In order to do so
we make use of the correspondence between stochastic Poisson brackets and 
quantum commutators \cite{Marra}. To achieve this correspondence
it is important that $\rho $ and $S$, as defined above, be canonical 
variables, in which case for a pair of dynamical variables $A$ and $B$, 
functionals of $\rho $ and $S$

\begin{equation}
\{ A, B\} _{s} = \int \left( \frac{\delta A}{\delta \rho } 
\frac{\delta B}{\delta S} -\frac{\delta A}{\delta S } 
\frac{\delta B}{\delta \rho }\right) dx \; .
\end{equation}

Guerra e Marra \cite{Marra} established the following correspondence

\begin{equation}
\{ A,B \} _{s}  = \imath  < [ A_q , B_q ]> \; ,
\end{equation} 

where $ A_q $ and $B_q $ are quantum operators and $A$ and $B$ are 
classical variables. The average value on the right hand side of (43) 
is defined as

\begin{equation}
<[A_q ,B_q ]> = \int \psi ^{*} [A_q ,B_q ] \psi \, dx \; .
\end{equation}

This result can be easily extended to our case ($dx = \prod _{i,m} 
d q_{m}^{i} $) and allows us to examine the issue of Lorentz 
invariance by computing the Poisson brackets for the generators of 
the Lorentz group now formulated in stochastic language. With 

\begin{equation}
M^{\mu \nu } = \int d\sigma (x^{\mu } P^{\nu } - x^{\nu } P^{\mu } ) 
\end{equation}

the critical element of the algebra is 

\begin{equation}
\{ M^{i-} , M^{j-} \} = i <[ M^{i-}_{q} , M^{j-}_{q} ]> \; .
\end{equation}

In order to close the algebra at a classical, but stochastical level, 
it is required that $D=26$, in agreement with the well known results 
in string theory \cite{Green}.

\subsection{Correlation Function}

In order to evaluate the correctness of our results, 
we calculate next the two-point correlation function using classical 
stochastic methods. We restrict the calculation to the non-zero 
normal modes in order to avoid the well-known infrared divergence 
\cite{Green}. For the ground state we have 

\begin{equation}
<x^{i} (\tau ) x^{i} (\tau ')> = \int  x^{i}\  
p(x^{i},\tau , x'^{i} , \tau ' )\ x'^{i}\   
\rho _{0} ( x'^{i}  )\ d x^{i} \ d x'^{i}  \; ,
\end{equation}

which gives

\begin{equation}
<x^{i} (\tau ) x^{i} (\tau ')> =  (D - 2)\ 2 \alpha ' 
\sum _{n \neq 0}  \frac{1}{n} \ 
\frac{\exp (n \tau ' )}{\exp (n \tau )} 
\end{equation}

This is the same result obtained from the standard first quantized 
string, as can be seen in ref.[1], if we continue $\tau $ to the Euclidean 
domain.

\section{Conclusion}

We have shown in this paper that the classical stochastic motion of an 
open bosonic string leads to a wave functional equation that 
correctly describes the first quantized string  and that for the consistency 
of our results requires the string to live in a 26 dimensional world. 
The agreement of our two-point correlation function calculated 
by classical stochastic methods, with the one 
obtained by the standard treatment, reinforces 
the soundness of our formulation.
Moreover, our results exhibit the naturalness of the 
string constant in playing 
the role of the diffusion constant at such small scales.

\acknowledgments
The authors acknowledge the support of the Brazilian 
Research Council, CNPq.

\end{document}